\documentclass[useAMS,usenatbib]{mn2e}

\usepackage{graphics}
\usepackage{epsfig}

\title[Globular Cluster Systems and Hot Gas]{
The Baryonic Halos of Elliptical Galaxies: Radial Distribution of 
Globular Clusters and Diffuse Hot Gas}
\author[D. A. Forbes et al.]{Duncan A. Forbes$^{1}$\thanks{E-mail:
dforbes@swin.edu.au}, Trevor Ponman$^{2}$ and Ewan O'Sullivan$^{2,3}$, 
\\
$^{1}$Centre for Astrophysics \& Supercomputing, Swinburne
University, 
Hawthorn VIC 3122, Australia\\
$^{2}$School of Physics and Astronomy, University of Birmingham, Edgbaston, Birmingham 
B15 2TT\\
$^{3}$Harvard-Smithsonian Center for Astrophysics, 60 Garden Street, Cambridge,
MA 02138, USA
}
\begin{document}


\pagerange{\pageref{firstpage}--\pageref{lastpage}} \pubyear{2002}

\maketitle

\label{firstpage}

\begin{abstract}
For a sample of 9 well-studied giant ellipticals we compare the
projected radial distribution of their red and blue globular
cluster (GC) subpopulations with their host galaxy 
stellar and X-ray surface brightness profiles. 
We support previous findings that the surface 
density distribution of {\it red} (metal-rich)
GCs follows that of the host galaxy starlight. 
We find good agreement between the outer slope of the {\it blue} GC 
surface density and that of the galaxy X-ray emission. 
This coincidence of projected radial profiles is likely due to the fact
that both blue GCs and X-ray emitting hot gas share the same
gravitational potential in equilibrium. When
deprojected the X-ray emitting hot gas has a radial
density dependence that is the square root of that for the GC density.  
We further show that the energy per unit mass for blue GCs is
roughly half that of the hot gas.

\end{abstract}


\section{Introduction}

The globular cluster (GC) systems of most, well-studied large galaxies reveal 
evidence for two subpopulations in colour (and by proxy in metallicity). 
Soon after the discovery of two subpopulations, it was shown that they have 
different spatial distributions with the {\it red} (metal-rich) subpopulation 
being more centrally concentrated than the more extended {\it
blue} (metal-poor) 
one (see Brodie \& Strader 2006 and references therein). 
Furthermore, the mean colours of both subpopulations 
correlate with the mass of the host galaxy, although the relation for the 
blue GCs is shallower than that for the red ones (Forbes et
al. 1997; Larsen et al. 2001; Peng et al. 2006). 

The {\it red} GCs are known to share a similar 1D radial, and 2D
spatial, distribution with that 
of the galaxy starlight in elliptical galaxies (Lee
et al. 1998;  Forbes et al. 2004; Dirsch et al. 2005; Lee et
al. 2008; Bassino et al. 2008; Faifer et
al. 2011; Strader et al. 2011). They also show similar kinematics 
(Norris et al. 2006, 2008; Lee et al. 2008; Pota et al. 2012) 
and stellar populations (Forbes \& Forte
2001; Norris et al. 2006; Spitler 2010; Forbes et al. 2011) in many cases
(although some exceptions do exist, e.g. Foster et al. 2011). This 
suggests that the red metal-rich GCs share a common formation history with 
the stellar component of an elliptical galaxy.

The {\it blue} GCs have been associated with elliptical galaxy halos in the
literature but the evidence for this is usually more indirect. 
Their radial surface density profiles are flatter than those 
of the starlight and they generally extend beyond the radius that can 
be accurately measured for the starlight. Forte et
al. (2007, 2011) have suggested that the `halo' component of elliptical
galaxy starlight is associated with the blue GC subpopulation. 
The halos of large elliptical galaxies also contain diffuse X-ray emitting 
hot gas. This gas is generally thought to be in
hydrostatic equilibrium, although the inner regions may
experience strong cooling via bremsstrahlung and line emission, and heating 
from AGN (e.g. O'Sullivan et al. 2003). 

In a cosmological context, 
rare over-dense fluctuations in the early Universe collapse
first. Thus the old stellar populations in a galaxy (e.g. the
halo field stars and blue metal-poor GCs) are expected
to share the same spatial distribution (Moore et al. 2006) and have the same
origin (Boley et al. 2009). 
Recent hydro-dynamical simulations have found similar
power-law slopes for the {\it stellar} halos of Milky Way like 
disk galaxies (e.g --2.6 to --3.4; Font et al. 2011) 
but lack the resolution to simulate GCs directly and have not yet
produced realistic X-ray surface brightness profiles. 
In the Milky Way,
both the halo field stars and the metal-poor GCs do share the same
3D spatial distribution which has a radial power-law slope of
$\sim$ --3.5 (Helmi 2008). Furthermore, it is now recognised
that a significant fraction of the halo field stars come from 
disrupted GCs  (Forbes \& Bridges 2010; Martell \& Grebel 2010).

As both the blue GCs and  the diffuse hot gas occupy 
the same halo gravitational potential they may reveal the same
radial distribution. Based on blue 
GC counts from a wide-field imaging study of Dirsch et
al. (2003) and the X-ray emission detected by the ROSAT satellite, Forte
et al. (2005) reported a good coincidence between the projected
surface density of blue GCs and the X-ray surface brightness
profile out to $\sim$110 kpc 
for the central dominant galaxy in the Fornax cluster 
NGC 1399.  As far as we are aware this is only case in the
literature of a direct comparison between the spatial
distribution of blue GCs and X-ray emission. It is therefore important to
understand whether NGC 1399 is a special case or whether such
similarities in the projected spatial distributions are common. 
A hint that the latter is true comes from the work of 
McLaughlin (1999) who pointed out the similarity in the {\it
total} GC system surface density profile 
with the combined stellar and hot gas radial
profiles for M87, NGC 1399 and NGC 4472. 
In the halo of a galaxy, these profiles are dominated by
the blue GCs and hot gas respectively. 

Lee et al. (2010) studied the large-scale distribution of GC
candidates through out the Virgo cluster using Sloan Digital Sky Survey
data. They concluded that this distribution was
qualitatively similar to that of the large-scale X-ray emission;
with the blue GCs more closely resembling the X-ray emission than
the red ones. Finally, we note that the global X-ray emission
from giant elliptical galaxies correlates better with the
velocity dispersion of the blue rather 
than the red GC subpopulation (Lee et al. 2008; Pota et al. 2012).

In this paper, we have collected data from the literature for a
small sample of giant elliptical galaxies that have 
diffuse hot gas halos and radially extended GC systems that can
be clearly separated into blue and red subpopulations.  
We explore the radial distributions of these baryonic
halo tracers and briefly discuss their interpretation. 

\section{Elliptical Galaxy Sample}

Our sample consists of two central cluster galaxies (M87 and NGC
1399), six 
giant ellipticals in groups or clusters and one isolated field
galaxy as listed in Table 1. 
Most are the central brightest galaxies in their host group/cluster
(i.e. BGG/BCG). 
Table 1 includes the type of galaxy, its environment, hot gas
temperature, distance, K-band luminosity, stellar velocity
dispersion, physical scale of 1 arcmin and the source of the
galaxy surface brightness profile. 
Apart from the field galaxy NGC 720, all are massive galaxies with M$_K$ $\sim$ --25 (M$_V$
$\sim$ --22) that dominate their group or subgroup within a
cluster, i.e. lie at the centre of their local potential well  and
X-ray emission.  In the case of the cD galaxy M87, it lies at the centre of
Virgo cluster potential well whereas NGC 4472 is the brightest
galaxy in the Virgo cluster dominating its own subgroup. 
NGC 1407 is the brightest galaxy in the massive 
Eridanus group (Romanowsky et al. 2009). NGC 4365 lies in the
W$'$ cloud, 
a group lying some 5 Mpc behind the Virgo
cluster. NGC 4636 lies in 
an X-ray distinct subgroup to the South of the main Virgo cluster. 
NGC 4649 is a giant Virgo elliptical galaxy. 
NGC 5846 is the brightest galaxy in the NGC 5846 group.

Our sample of elliptical galaxies was  
selected, from the literature, to have extended radial profiles for both GCs and
diffuse X-ray emission. In particular, the blue GC density profile was
required to extend to at least 8 arcmin (about 45 kpc at the
typical distance of our sample). Further details of the
selection process and literature data are discussed below. 

\begin{table*}
\caption{Elliptical galaxy sample. Type, Environment, Distance
and K-band magnitude are from NASA Extragalactic
Database. X-ray temperatures are from Diehl \& Statler (2008). 
Stellar velocity dispersions ($\sigma_{\ast}$) are from HyperLeda; Surface 
Brightness profile. References for surface brightness (SB) profiles are: (1) Kormendy et 
al. (2009); (2) Goudrooij et al. (1994); (3) Forte et al. (2005); (4) Spitler et al. 
(2012); (5) Blom, Spitler \& Forbes (2012); (6) Caon et al. (1994); (7) MacDonald et al. 
(2011); (8) Kronawitter et al. (2000). 
}
\begin{tabular}{lcccccccc}
\hline
Galaxy & Type & Envir. & T$_X$ & Dist. & M$_K$ & $\sigma_{\ast}$ & kpc/$'$ & SB profile\\
 & & & (keV) & (Mpc) & (mag) & (km/s) & &\\
\hline
M87 & cD & Virgo  & 2.50 & 16.7 & --25.29 & 334 & 4.9 & (1)\\
NGC 720 & gE & Field & 0.57 & 24.0 & --24.63 & 241 & 7.0 & (2)\\
NGC 1399 & BCG & Fornax  & 1.13 & 19.0 & --25.06 & 346 & 5.5 & (3)\\
NGC 1407 & BGG & Group  & 0.87 & 23.8 & --25.15 & 271 & 6.9 & (4)\\
NGC 4365 & BGG & Group  & 0.64 & 21.4 & --25.00 & 256 & 6.2 & (5)\\
NGC 4472 & BCG & Virgo  & 0.97 & 16.1 & --25.61 & 294 & 4.7 & (6)\\
NGC 4636 & BGG & Group  & 0.69 & 15.9 & --24.57 & 203 & 4.6 & (6)\\
NGC 4649 & gE & Virgo  & 0.80 & 16.5 & --25.34 & 335 & 4.8 & (7)\\
NGC 5846 & BGG & Group  & 0.71 & 26.7 & --25.18 & 239 & 7.8 & (8)\\
\hline
\end{tabular}
\end{table*}


\begin{table*}
\caption{Globular cluster data. 
References for surface density and filter data 
are: (1) Strader et al. (2011); (2) Kartha et al. (2012, in prep)
(3) Bassino et al. (2006); (4) Spitler et al. (2012); (5) Blom,
Spitler \& Forbes (2012); 
(6) Rhode et al. (2011, priv. comm.); (7) Dirsch et
al. (2005); (8) Lee et al. (2008); (9) Pota et al. (2012). 
All blue and red GC velocity dispersions are from Pota et
al. (2012), except M87 from Lee et al. (2008). Total number of
globular clusters are from Ashman \& Zepf (1998), except for NGC 1407
from Perrett et al. (1997). 
}
\begin{tabular}{lccccc}
\hline
Galaxy & Density & Filters & $\sigma_{BGC}$ & $\sigma_{RGC}$ & N$_{GC}$\\
 & & & (km/s) & (km/s) & \\
\hline
M87 & (1) & gri & 414 & 380 & 13000\\
NGC 720 &  (2) & gi & -- & -- & 660\\
NGC 1399 & (3) & CR & 337 & 277 & 5410\\
NGC 1407 & (4) & gri & 239 & 212 & 2640\\
NGC 4365 & (5) & gri & 238 & 261 & 2500\\
NGC 4472 & (6) & BVR & 337 & 257 & 6300\\
NGC 4636 & (7) & CR & 236 & 194 & 3000\\
NGC 4649 & (8) & CT$_1$ & 197 & 218 & 5100\\
NGC 5846 & (9) & gri & 264 & 206 & 2200\\
\hline
\end{tabular}
\end{table*}

\section{X-ray Surface Brightness Profiles}

The X-ray emission of giant elliptical
galaxies is dominated by diffuse hot gas and their surface
brightness profiles tend to range from core-like to power-law in
their inner regions, with an outer power-law slope. Such profiles
are usually quantified by a so-called beta model, i.e.\\

$S(r) = S_0[1 + (r/r_c)^2]^{-3\beta_{X} + 0.5}$ \hspace{1.1in}(1)\\

\noindent
Here we take the beta model fit parameters (i.e. the core radius r$_c$
and the $\beta_X$ value) from the X-ray surface brightness profiles
in the literature. 

Although the Chandra satellite has superior resolution and
XMM-Newton has better sensitivity, the ROSAT satellite is
generally preferred for its extended radial coverage when studying the
outer halos of giant elliptical galaxies. For the two central
cluster galaxies (M87 and NGC 1399) 
in our sample we use beta model fits to the 
ROSAT X-ray surface brightness profiles. These fits 
extend to over 100 arcmin for M87 (Bohringer et al. 1994) 
and to 40 arcmin for NGC 1399 (Jones et al. 1997). For NGC 5846 a
deep (120 ksec) Chandra image is available from Machacek et
al. (2011). 

For the other giant ellipticals, X-ray profiles come from ROSAT data in
the large sample study of O'Sullivan, Ponman \& Collins
(2003). A detailed
description of the data reduction and analysis can be found in
O'Sullivan et al. Briefly, 
periods of high background (deviating by $>$2$\sigma$
from the mean event rate) were excluded from each dataset, point sources of
$>$4.5$\sigma$ significance were excluded (excepting those within the
D$_{25}$ radius of the target galaxy), and a background model determined
based on a large-radius annulus. A 0.5-2 keV image was extracted and
corrected for vignetting, and a flat background level was determined and
subtracted from the image. A standard beta model 
was fit to the data, after convolution
with the point spread function appropriate for the mean photon energy
of the source. 

All the sample galaxies, including M87 and NGC 1399, are 
highly X-ray luminous and gas-rich. We
therefore expect individual X-ray binaries to have little, or no impact, on
the surface brightness fits. In a Chandra X-ray study that
included our galaxies, Diehl \& Statler (2007) estimated an average
unresolved fraction of $<8\%$. 
A small number of bright point sources are visible
in the O'Sullivan et al. data, and they 
are excluded from the fit. The remaining unresolved
sources could potentially affect the fit in the galaxy core, but we are
primarily interested in the outer parts of each galaxy, where GCs 
can be reliably detected. Here, the density of X-ray binaries will
be low, and therefore very unlikely to have any significant influence on
the surface brightness modelling. We also note that when X-ray
binaries are located in GCs they are typically found in red
rather than blue GCs (Paolillo et al. 2011). 

We exclude all galaxies that have X-ray profiles classified as
`uncertain' (e.g. NGC 4552) or noted to have  
strong AGN activity (e.g. NGC 5128) by O'Sullivan et al. (2003).

\section{Globular Cluster Surface Densities}

The GC data used here come from a variety sources in the literature. The
original imaging data should be multi-filter (to clearly
distinguish the blue and red subpopulations) with a wide
field-of-view (to
cover a large radial range). 

The first 
criterion of using more than one filter is met by most GC studies of giant
ellipticals. Thus with sufficient numbers of GCs the system can
be separated into blue and red subpopulations based on a simple
colour division. Here we adopt the blue/red definitions from the
original literature studies. 

For the second criterion, our sample is restricted to those
galaxies which have
GC system surface density profiles out to at 
least 8 arcmin (about 45 kpc for our sample). The minimum radial
extent is for NGC 4649 which reaches to 8.75 arcmin using data
from Lee et
al. (2008). 
The typical effective
radius of the galaxy starlight for our sample is 5--10 kpc; thus
the GC data typically probe to several effective
radii in terms of the host galaxy starlight. Requiring large radial
coverage of nearby galaxies effectively restricts studies to
those that use a ground-based telescope. Our final sample of
giant elliptical galaxies is summarised in Table 1. 

All GC profiles have been corrected for background contamination
in the original work. However, the level of the background
contamination and the ability to subtract it accurately vary
for each dataset. This uncertainty is generally not captured in the
(Poisson) error bars quoted in the literature. 
Although good seeing conditions help for GC
studies, the introduction of a third filter can significantly
decrease background levels as GC candidates can be selected in
colour-colour space. For example, Romanowsky et al. (2009) found
a contamination rate of $\sim$5\% in follow-up spectroscopy of
GCs in NGC 1407 selected from three filters. In Table 2, along
with the source of the GC surface density data, we list whether it comes from 2
or 3 filter imaging. Thus the best quality GC data are available
for M87, NGC 1407, NGC 4365, NGC 4472 and NGC 5846. The total
number of objects in the GC system, from the literature, is also included. 

Traditionally, GC surface density profiles have been fit by a simple
power-law. However, this approach often fits the central regions
in which the density flattens off, perhaps due to destruction
effects (Miocchi et al. 2006) and
hence tends to underestimate the outer
slope. A better approach, which is becoming more common
(e.g. Strader et al. 2011; Blom, Spitler \& Forbes 2012) is to fit Sersic
(1968) profiles to the GC density data as has been done for
galaxy surface brightness profiles for a number of years.  
Here we have decided to 
fit the GC surface density data with the same beta profile form as
used in fits to the X-ray surface brightness data in order to
facilitate a direct comparison. Such beta profiles, like Sersic
ones, fit an inner core region and the changing slope of the
profile.



\section{Results}

\subsection{M87}

In Figure 1 we show a comparison between the GC system surface
density and the X-ray emission as quantified by a
beta model (Eq. 1) for the central galaxy in the Virgo cluster M87. 
Although not
optically the brightest in the cluster, it lies at the centre of
the Virgo cluster X-ray emission. 
In this figure, and subsequent ones, the X-ray profile has
been normalised to a value of 1 at 10 arcmin radius and the GCs
have been arbitrarily normalised in the vertical
axis. 

For the {\it blue} GCs we show the individual data points and a
beta model fit. We also show the X-ray beta model fit and a
$\pm$5\% variation (a typical value for well-constrained X-ray
profiles) on the $\beta_X$ slope.
The blue GCs and X-ray beta model for M87 reveal a similar
slope from a few arcminutes to 30 arcminutes. Only in the inner
regions, in which the X-ray emission 
may be affected by thermal heating asociated with
the AGN/jet, does the coincidence begin to break down. 
We note that Lee et al. (2010) measured a power-law
slope of --1.49 $\pm$ 0.09 for the blue GCs within 40 arcmin, and
that this compares well with our data which have a power-law
slope of --1.54 $\pm$
0.04.

We also
show the similarity between the {\it red} GC 
surface density profile and that of the galaxy
starlight (from Kormendy et al. 2009). 
Such a connection between the red GC subpopulation and
the starlight of the galaxy has been observed in many
galaxies as noted in the Introduction 
(see Brodie \& Strader 2006 for an overview). 

\subsection{NGC 1399}

In Figure 2 we show the GC, X-ray and stellar profiles for the central
galaxy of the Fornax cluster NGC 1399. 
Forte et al. (2005) highlighted the similar slopes, between 1 and
20 arcmin, of the blue GC data of Dirsch et al. (2003) 
and the X-ray surface brightness
profile of Jones et al. (1997). Here we use the same X-ray data (rather than 
the more recent Chandra data of Scharf et al. (2005) which only extend to less than 
2 arcmin) 
but show the more recent and more radially extended GC data of 
Bassino et al. (2006). The figure supports the claim of Forte et
al. but also shows that the agreement in profile 
slopes begins to break down for
radii beyond 20 arcmin. We note that the outer region GC data are
highly sensitive to contributions from other Fornax cluster
galaxies (e.g. NGC 1404; Bekki et al. 2003, Schuberth et al. 2011). We also show the red
GCs compared to the stellar surface brightness profile from Forte
et al. (2005), again with an arbitrary normalisation. Similar to
the situation in M87, the red GCs and the starlight show consistent
profile slopes. 

\subsection{NGC 5846}

In Figure 3 we show the GC, X-ray and stellar profiles for the
central group galaxy NGC 5846. The X-ray beta model fit, to the deep
(120 ksec) Chandra X-ray surface brightness profile, is from
Machacek et al. (2011). They note that the galaxy shows signs of
central AGN activity (e.g. X-ray cavities) and non-hydrostatic 
gas motions, yet the profile is fairly well represented by a
single beta model to 12 arcmin. As for M87 and NGC 1399, the red
GCs are well matched to the starlight profile, while the blue GCs
are consistent with the X-ray profile over most radii (the
innermost regions may be affected by GC destruction processes;
Miocchi et al. 2006). 

\subsection{The O'Sullivan et al. sample}

In Figures 4 and 5 we show galaxies from the giant elliptical galaxy 
sample of O'Sullivan
et al. (2003) for which we were able to find 
GC surface density data that extended to at least 8 arcmin
(i.e. well beyond any GC core region). 
The figures show the {\it blue} GC surface density data from
the literature, beta model fits to those data, and 
beta model fits (and their uncertainty) to the ROSAT X-ray surface
brightness profiles as determined by O'Sullivan. 
For most galaxies the surface density of blue GCs in the outer regions is well-matched to the
X-ray profile. The main exception is NGC 4365 which reveals a GC
profile that is significantly flatter than the X-ray one. This
galaxy is currently undergoing an interaction (Mihos et al. 
2012) and it is possible that
the accreted galaxy has contributed extra GCs to the outer
regions ($>$ 10 arcmin) of NGC 4365. 

All of the GC systems in Figures 4 and 5 reveal a
flattening (i.e. towards a constant density) at small radii. This
has been observed in many GC systems (Forbes et al. 1996) and may
be due to GC destruction processes, such as bulge shocking
(Miocchi et al. 2006). 

\begin{table*}
\caption{Beta slope values for X-ray and globular cluster radial profiles. References 
for X-ray data are: (1) Bohringer et al. (1994); (2) O'Sullivan et al. (2003); (3) Jones 
et al. (1997); (4) Machacek et al. (2011). 
}
\begin{tabular}{lcccc}
\hline
Galaxy & X-ray & $\beta_X$ & $\beta_{BGC}$ & $\beta_{RGC}$\\
\hline
M87 & (1) & 0.47 & 0.43 $\pm$ 0.13 & 0.49 $\pm$ 0.01\\ 
NGC 720 & (2) & 0.483 $\pm$ 0.01 & 0.60 $\pm$ 0.09 & 0.72 $\pm$ 0.12\\
NGC 1399 & (3) & 0.35 & 0.42 $\pm$ 0.05 & 0.51 $\pm$ 0.02\\
NGC 1407 & (2) & 0.56$^{+0.02}_{-0.01}$ & 0.58 $\pm$ 0.17 & 0.86
$\pm$ 0.04\\
NGC 4365 & (2) & 0.60$^{+0.04}_{-0.03}$ & 0.46 $\pm$ 0.12 & 0.61
$\pm$ 0.03\\
NGC 4472 & (2) & 0.597$^{+0.009}_{-0.008}$ & 0.67 $\pm$ 0.22 &
0.52 $\pm$ 0.02\\ 
NGC 4636 & (2) & 0.535$^{+0.007}_{-0.006}$ & 0.69 $\pm$ 0.09 &
0.53 $\pm$ 0.02\\
NGC 4649 & (2) & 0.567$\pm{0.08}$ & 0.55 $\pm$ 0.09 & 0.63 $\pm$ 0.11\\
NGC 5846 & (4) & 0.45 & 0.51 $\pm$ 0.10 & 0.54
$\pm$ 0.09\\
\hline
Mean & -- & 0.51 $\pm$ 0.03 & 0.54 $\pm$ 0.03 & 0.60 $\pm$ 0.04\\
\hline
\end{tabular}
\end{table*}

As mentioned earlier, we have fit the blue GC surface density
data with a beta model (Eq. 1). For NGC 4636 
we were unable to obtain stable fits to the error-weighted
data and so we chose to fit with equal weighting. The GC data for
NGC 4636 show an increased density at large radii. This is
likely an indication of the uncertainty in the background
subtraction. If we fit only the blue GC data interior to 12.5 arcmin we
derive the same beta slope within the
errors quoted in Table 3.
For NGC 4649
the limited radial extent did not allow for a beta model fit for
the blue GCs. In this case we fit a simple
power-law fit to the data beyond 4 arcmin which gives an equivalent
beta slope of 0.55 $\pm$ 0.09 and is quoted in Table 3. 
The results of our beta model fitting to the blue GC density profiles are
given in Table 3, along with the $\beta_X$ values from the X-ray
surface brightness fits from the literature. Table 3 also gives
mean values and the error on the mean. The  
X-ray and blue GC beta slopes are consistent within the errors,
while the red GCs have higher beta values than the X-ray profiles
at the $\sim$2$\sigma$ level. 

Figures 4 and 5 also show the {\it red} GC data with the galaxy surface
brightness profile from the literature. 
The galaxy profiles are from Goudfrooij et al. (1994)
for NGC 720; Spitler et al. (2012) for
NGC 1407; Blom, Spitler \& Forbes (2012) for NGC 4365; Caon et al. (1994) for
NGC 4472 and NGC 4636 and MacDonald et al. (2011) for NGC 4649.
Although the galaxy
starlight does not extend as far as the GC data (particularly for
NGC 720), the GCs and the starlight show
similar slopes in their outer regions. Given that the blue and
red GCs occupy the same gravitational potential but have
different density profiles, the Jeans equation suggests that they also have
different orbital anisotropy properties. 


\subsection{Summary}

We support previous findings that the {\it red} GC
density profiles generally follow the starlight in elliptical
galaxies.
Most elliptical galaxies in the sample show good
coincidence between the outer slope of the {\it blue} GC profile
and the X-ray surface brightness. 
We note that the measured GC slopes 
are dependent on the level of background contamination
and hence the overall quality of the data which is not captured
by the purely Poisson errors given in the literature. 
The average slopes of the blue GCs and X-ray emission are
consistent within the errors, although our sample is small.

\section{Discussion}

In the previous section we showed that the outer slope of the X-ray
emission
(with a mean slope of $\beta_X$ = 0.51 $\pm$ 0.03) was in good
agreement, in most cases, with
the blue GCs surface density slope (mean $\beta_{BGC}$ = 0.54 $\pm$ 0.03).
Could this similarity between the projected blue GC and X-ray profiles
indicate a direct physical connection between the metal-poor
GC system and the diffuse hot gas in an
elliptical galaxy halo? One possibility is that GCs have formed within
hot gas that permeates the halos of giant ellipticals.
For example, in the model of Fall \&
Rees (1985) the hot gas acts to compress colder gas
clouds. For metal-poor gas this gives rise to
a characteristic mass of $\sim$10$^6$ M${_\odot}$ which is
typical of GCs. However a problem with this scenario is that
metal-poor GCs also form in dwarf galaxies which lack
hot gas halos.

More probably, the coincidence in profile shape between blue GCs and
X-ray emission from hot gas reflects the fact that for our giant
elliptical
galaxy sample both
halo tracers are in equilibrium within the potential
well centred on the host galaxy.
The observed correlations of hot gas X-ray emission
and the blue GC velocity dispersion (Lee et al. 2008; Pota et
al. 2012) further supports this interpretation.

If this is the case, then
the similar outer slopes seen for the blue GCs and the X-ray emission
has an interesting implication for the relative
density profiles of the two tracers and their specific energies.
The X-ray emissivity, which projects to give surface brightness,
is proportional to the
square of the hot gas density, whereas the
projected GC surface density scales linearly with the GC
density. So similar outer slopes in projection imply
that the hot gas density scales with the
GC density to the power of 0.5. In other words, the GCs have a
steeper 3D radial density distribution than the hot gas.

The radial distribution of a tracer in equilibrium within a
gravitational potential well is related to its specific energy, and
hence the relative slopes of the gas and GC profiles will depend
on the ratios of their specific energies, traditionally
denoted $\beta_{spec}$ in the cluster literature. In our case:\\

\noindent
$\beta_{spec} = 0.5(3\sigma_{BGC}^2) ~/~ (3kT_X/2m)$, \hspace{1.1in} (2)\\

\noindent
where $\sigma_{BGC}$ is the blue GC velocity dispersion, $T_X$
is the hot gas temperature, $k$ is the Boltzman constant and
$m$ is the mean particle mass of the hot gas.
Using $m =0.6$ amu, this reduces to:\\

\noindent
$\beta_{spec} =(\sigma_{BGC} ~/~ 1000\,{\rm km~s^{-1}})^2 ~(6.22\,{\rm
keV}/T_X)$.
\hspace{0.5in} (3)\\

\noindent
Taking X-ray temperatures (T$_X$) from Table 1 and
blue velocity dispersions ($\sigma_{BGC}$)
from Table 2, the resulting $\beta_{spec}$
values have a mean of 0.52 for our sample. In other words, the
blue GCs have about half the specific energy of the hot gas.

Assuming that the GCs are virialised and follow the Jeans
equation, and the hot gas is in hydrostatic
equilibrium, the value of $\beta_{spec}$ is related to the gradients in
the density and temperature/velocity dispersion of the two tracers
by equation 35 from Bahcall \& Lubin (1994).

Although kinematic studies of GCs are limited,
to first order blue GCs reveal flat velocity dispersion profiles
(e.g. Lee et al. 2008; Pota et al. 2012). The orbital anisotropy
for blue GCs is not well determined but orbits appear to be close to
isotropic on average. Similarly, X-ray temperature profiles,
beyond the very inner regions, are
close to flat on average (O'Sullivan et al. 2003).
For the case of isotropic GC orbits with flat velocity dispersion
gradient and isothermal gas, the equation simplifies such that:\\

\noindent
$\beta_{spec} = \frac{d ln \rho_{gas} ~/~ d
ln r}{d ln \rho_{BGC} ~/~ d ln r }$ \hspace{1.9in} (4)\\

\noindent
where $\rho_{gas}$ and $\rho_{BGC}$ are the density of hot gas
and blue GCs respectively.

Hence, since we find $\beta_{spec}$ $\sim$ 0.5, it follows that we would
expect the logarithmic slope of the gas density to be one half of that of
the blue GCs, and hence the X-ray emissivity (scaling as density squared)
would indeed have the same slope as the density of blue GCs.

For a beta model (Eq. 1) the logarithmic slope at large radii is given
by $-6 \beta + 1$. Given the mean beta value for the blue GCs of
$\beta_{BGC} = 0.54$ this corresponds to a projected outer slope of -2.24.
If the GCs have spherical symmetry then
the {\it deprojected} 3D slope for the blue GCs will be
-3.24 (in comparison the surface density of blue GCs in the Milky
Way has a 3D slope of --3.5; Helmi 2008).
In comparison, the projected X-ray emissivity (with mean $\beta_X$=0.51
from Table~3) has a slope of -2.06, and hence a 3D slope of -3.06, so
that the average 3D gas density slope is -1.53, roughly one half
that of the blue GCs.

This result is reminiscent of that found by Osmond \& Ponman
(2004) for galaxy groups. They derived the specific energy of the
galaxies and compared it to that for the hot gas in galaxy groups.
The groups had a mean
$\beta_{spec}$ = 0.75 indicating that the galaxies had a lower
specific energy than the hot gas within the group. (We note that
galaxy clusters have $\beta_{spec}$ $\sim$
1.) Here, we find that the blue GCs have a specific energy
($\beta_{spec} \sim 0.5$) lower
than that of galaxies in groups
and much lower than the expected value for the group potential.
Thus whatever process gives rise to blue GCs in the halos of
giant elliptical galaxies must deposit them with a relatively low
specific energy. 
If blue GCs have been accreted relatively recently then the likely source
would be more massive galaxies which, unlike dwarfs, are able to lose specific
energy on a Gyr timescale through dynamical friction. However,
perhaps more likely is the accretion and disruption of dwarf
galaxies at early times when the group/cluster was less massive
than today.

\section{Summary and Conclusions}

For our sample of 9 giant ellipticals we have shown that the 1D
radial surface density of {\it red} GCs, beyond the inner
regions, is well matched to the stellar surface brightness
profile. Thus, combining with literature studies,
{\it red} GCs have the following properties:\\
$\bullet$ similar 1D radial distribution to the host galaxy stars\\
$\bullet$ similar 2D spatial distribution to the host galaxy stars\\
$\bullet$ similar stellar populations to the host galaxy stars\\
$\bullet$ correlate with galaxy stellar mass and stellar
velocity dispersion\\

The {\it blue} GCs in the Milky Way share many properties with
the stellar halo. For elliptical galaxies the connection between
the {\it blue} GCs and the hot gas halo was based on the
similarity in 1D radial profiles for one galaxy,
NGC 1399 (Forte et al. 2005). Here we have revisited NGC 1399 and included
an additional 8 giant ellipticals, finding that in most cases
there is a good coincidence between the outer slopes of the {\it
blue} GC surface density and the X-ray surface brightness. Thus
we provide evidence for a connection between {\it blue} GCs and
the halos of elliptical galaxies that until now was largely
conjecture. This finding means we can use {\it blue} GCs as a
halo tracer for lower luminosity ellipticals for which the X-ray
emission is undetected.

As the X-ray emission scales with gas density squared and blue GC
surface density scales linearly with GC density, the
similarity in {\it projected} outer slopes
implies that the 3D density distribution of the hot gas scales
with the blue GC density to the power of a half.
Consistent with this, we calculate that the
specific energy of the blue GCs is about one half
that of the hot gas.


\begin{figure*}
\begin{center}
\includegraphics[angle=0,scale=0.7]{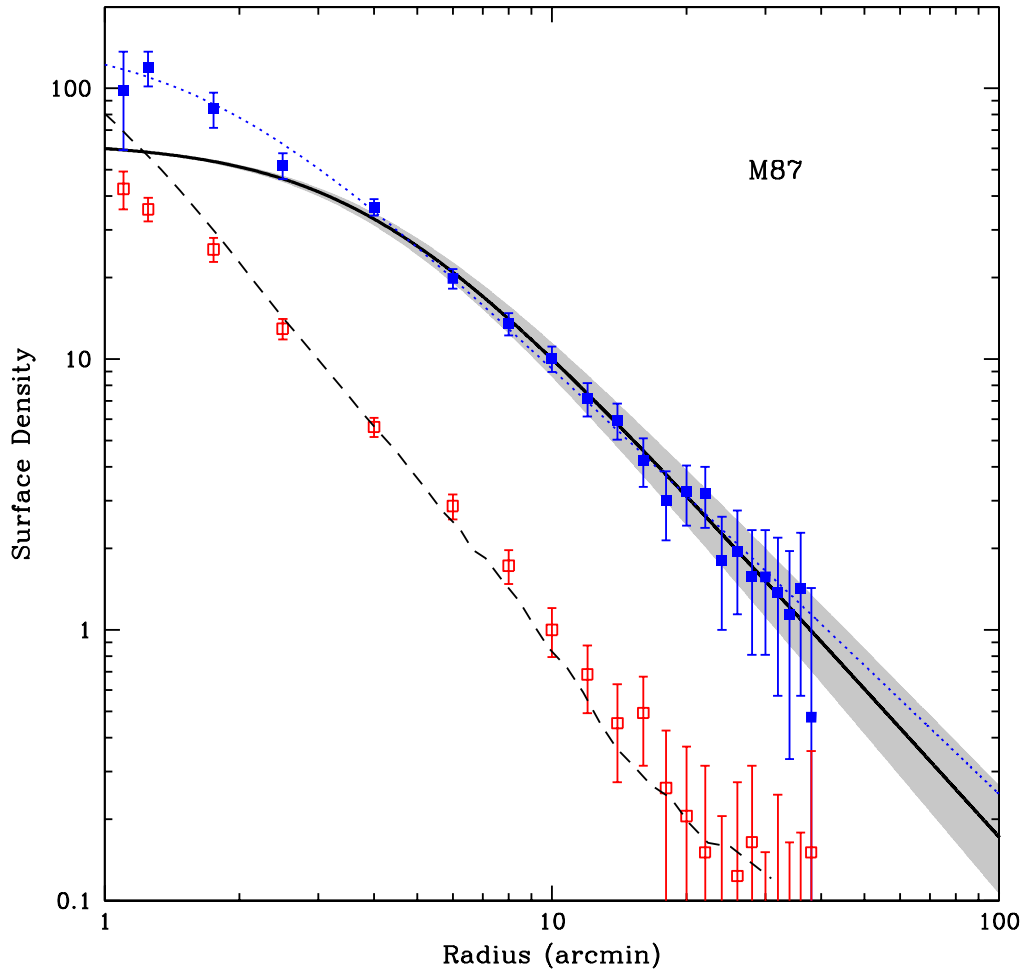}
\caption{Globular cluster and X-ray surface density profiles for 
the central dominant Virgo galaxy M87. The blue 
(blue filled squares) and red (red open squares) globular cluster
system density profiles are shown along with a beta model fit to
the blue GC data (dotted line). The X-ray surface brightness
from a beta model fit to beyond 100 arcmin (solid line) with a $\pm$5\% slope
uncertainty (shaded region) and
the galaxy starlight profile from Kormendy et al. (2009) 
(dashed line) are also shown.
All profiles have an arbitrary vertical normalisation. 
The red GCs follow the galaxy stellar component while the blue GCs
trace the X-ray profile. 
}
\end{center}
\end{figure*}

\begin{figure*}
\begin{center}
\includegraphics[angle=0,scale=0.7]{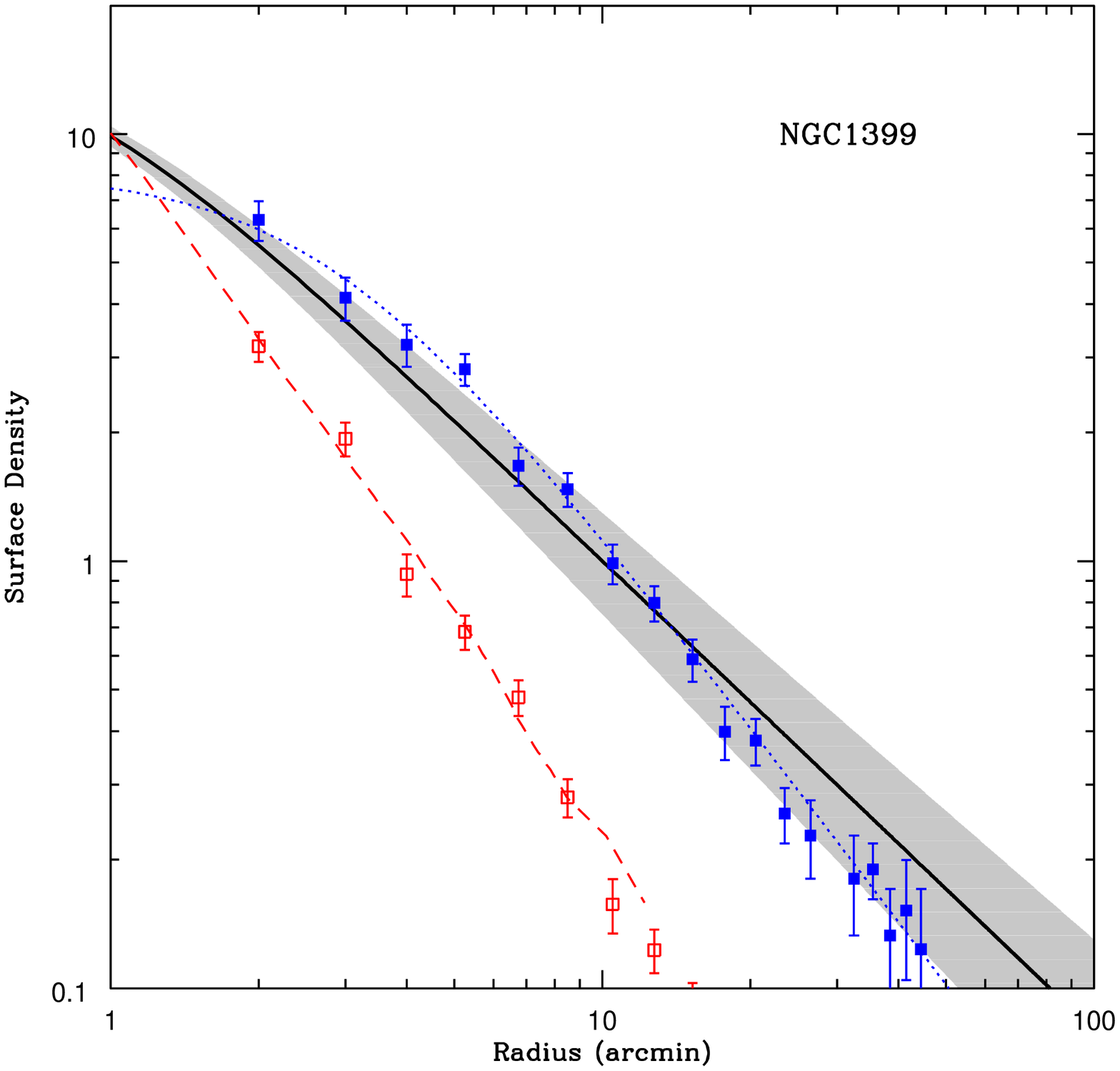}
\caption{Globular cluster and X-ray surface density profiles for 
the central dominant Fornax galaxy NGC 1399. The blue
(blue filled squares) globular cluster
density profile is shown along with a beta model fit to 
the blue GC data (dotted line). The X-ray surface brightness
from a beta model fit to radii $\le$ 40 arcmin (solid line) with a $\pm$5\%
slope uncertainty (shaded region) is
shown. We also show the red (red filled squares) globular cluster
density profile along with the galaxy starlight 
profile from Forte et al. (2005) (long dashed line). 
All profiles have an arbitrary vertical normalisation. 
The red GCs follow the galaxy stellar component while the blue GCs
have a similar slope to the X-ray profile. 
}
\end{center}
\end{figure*}

\begin{figure*}
\begin{center}
\includegraphics[angle=0,scale=0.7]{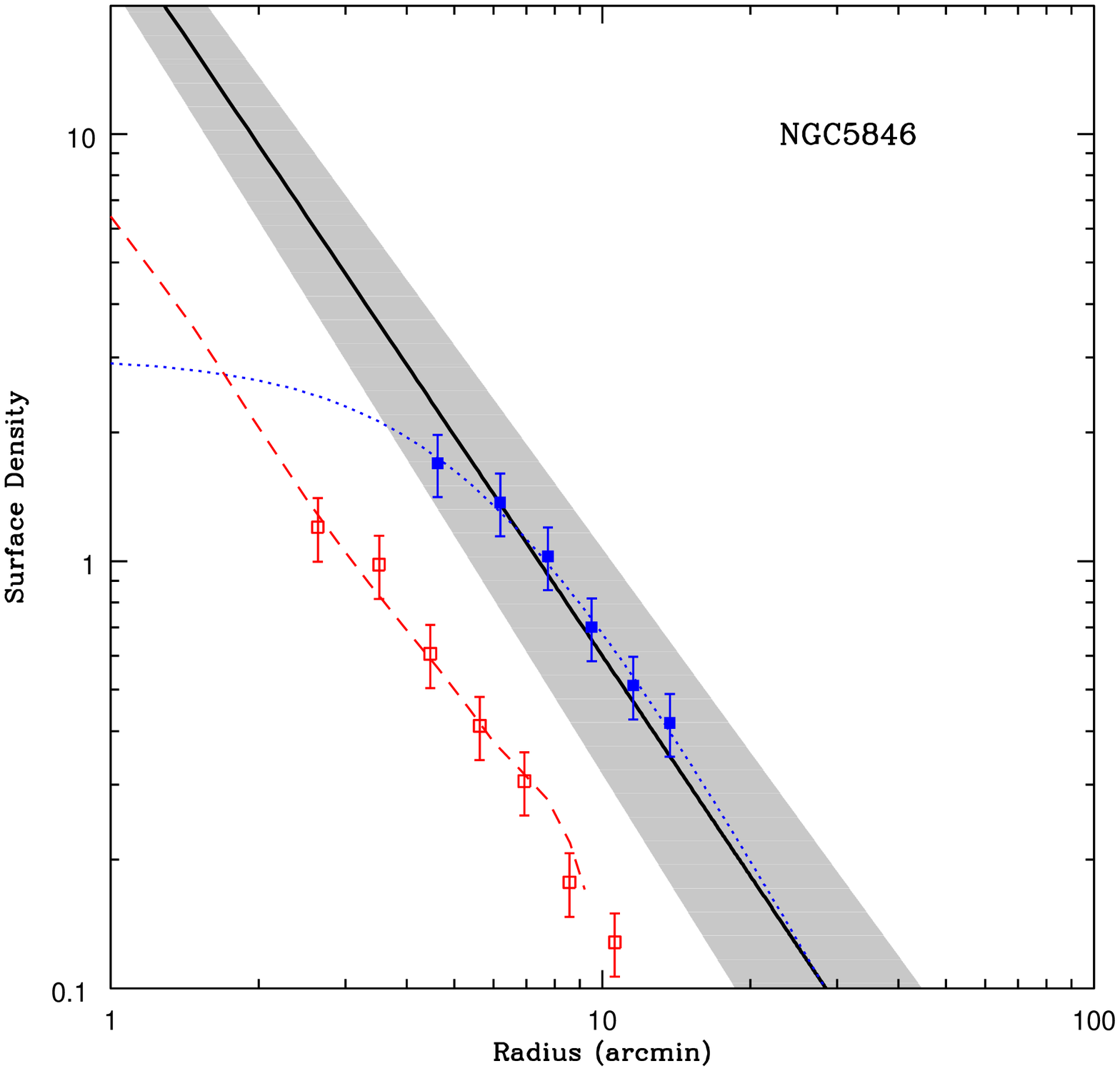}
\caption{Globular cluster and X-ray surface density profiles for 
the central group galaxy NGC 5846. The blue
(blue filled squares) globular cluster
density profile is shown along with a beta model fit to 
the blue GC data (dotted line). The X-ray surface brightness
from a beta model fit to radii $\le$ 12 arcmin (solid line) with a $\pm$5\%
slope uncertainty (shaded region) is
shown. We also show the red (red filled squares) globular cluster
density profile along with the galaxy starlight 
profile from Kronawitter et al. (2000) (long dashed line). 
All profiles have an arbitrary vertical normalisation. 
The red GCs follow the galaxy stellar component while the blue GCs
have a similar slope to the X-ray profile. 
}
\end{center}
\end{figure*}


\begin{figure*}
\begin{center}
\includegraphics[angle=-90,scale=0.7]{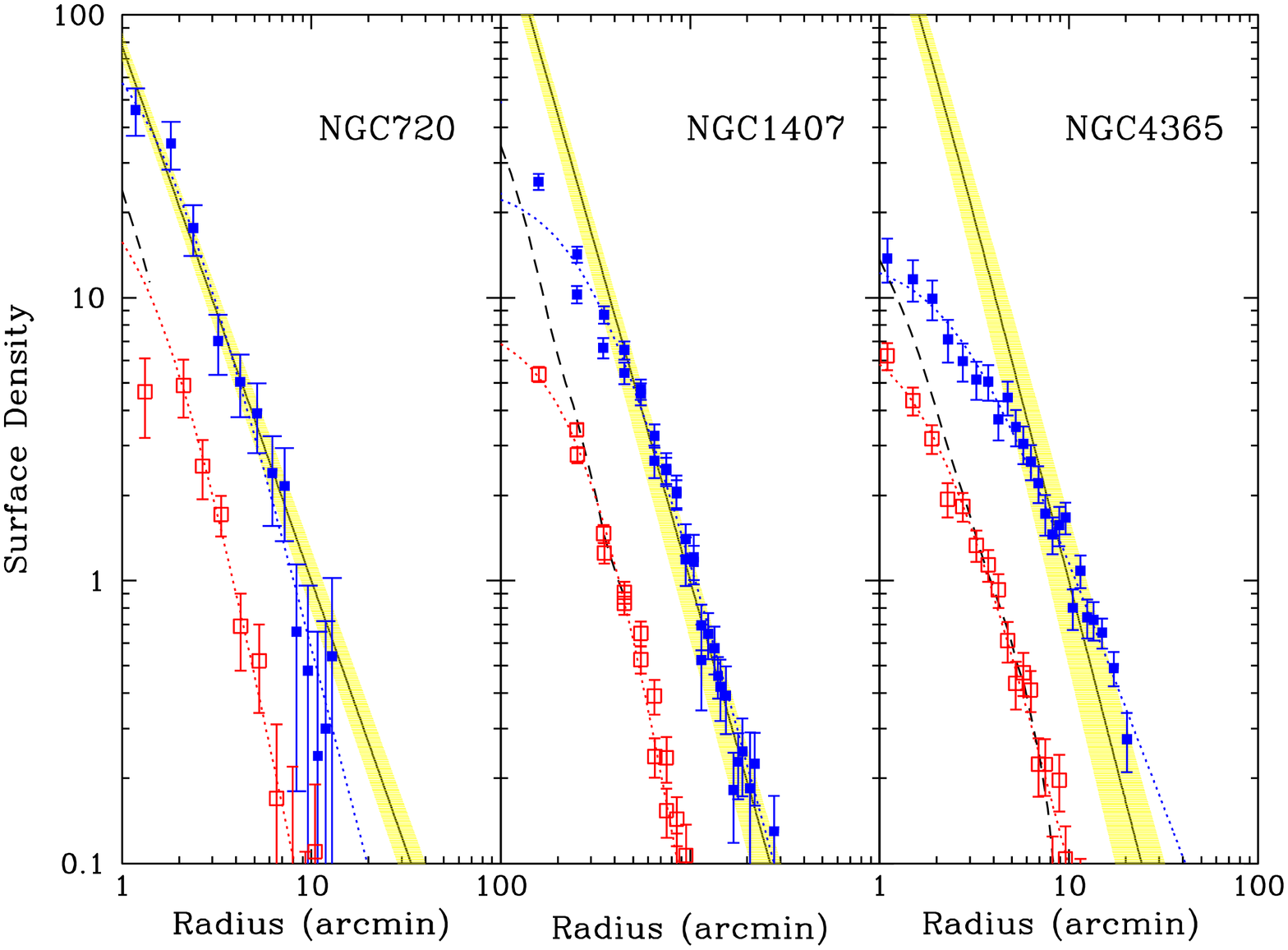}
\caption{Globular cluster and extrapolated X-ray surface density profiles for
  NGC 720, 1407 and 4365. The X-ray surface brightness beta
  model fits and uncertainty (line and yellow shaded region) 
are all taken from the survey of 
O`Sullivan et al. (2003). 
The blue 
(blue filled squares) and red (red open squares) globular cluster
density profiles are shown along with a beta model fits (blue and
red dotted lines). The galaxy surface brightness profile is shown
by a white dashed line.
All profiles
  have an arbitrary vertical normalisation. For most galaxies,
  the outer slopes of the X-ray and blue GC profiles are similar,
  as are the galaxy surface brightness and red GC profiles.
}
\end{center}
\end{figure*}

\begin{figure*}
\begin{center}
\includegraphics[angle=-90,scale=0.7]{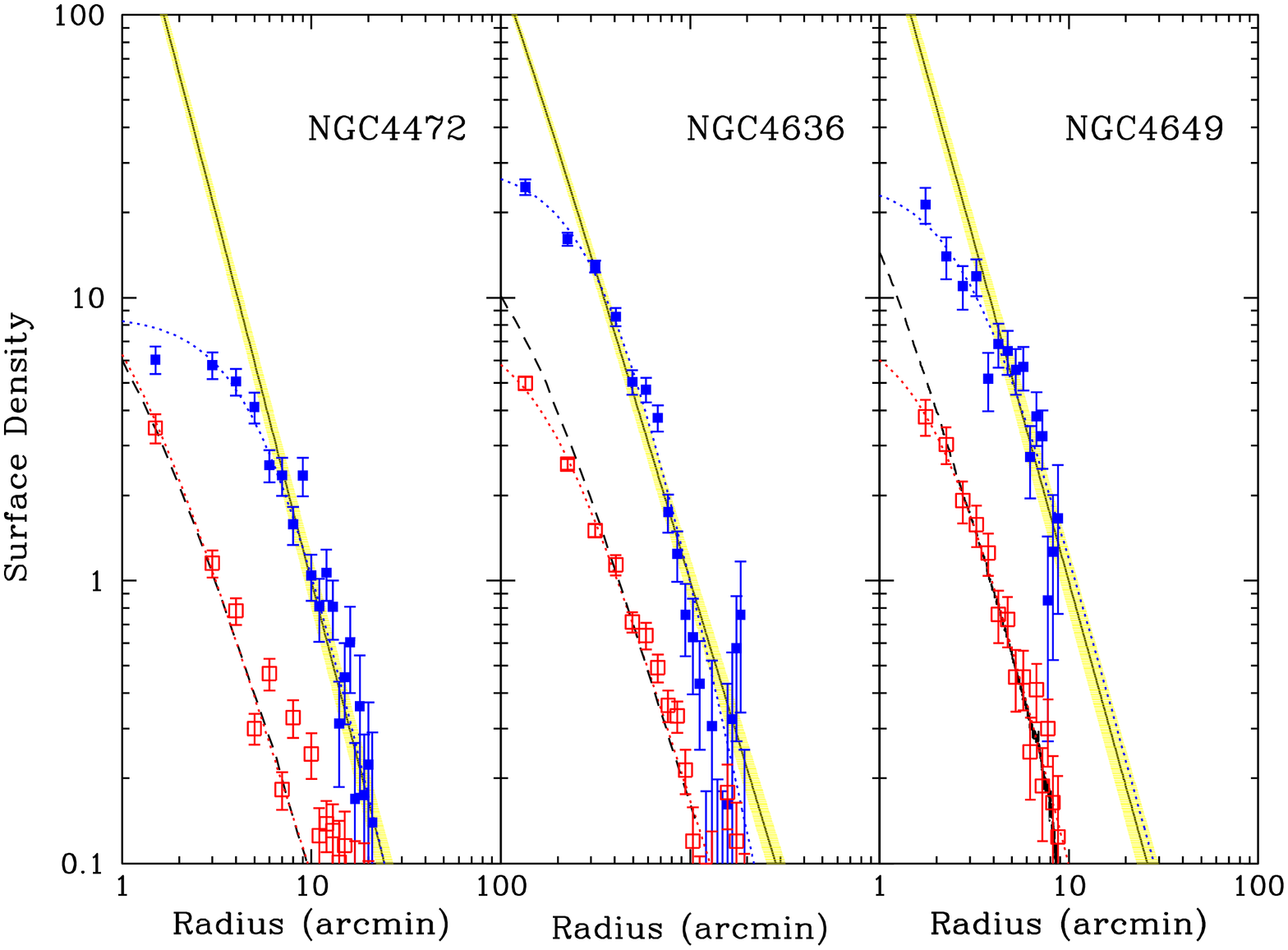}
\caption{Globular cluster and extrapolated X-ray surface density profiles for
  NGC 4472, 4636 and 4649. The X-ray surface brightness beta
  model fits and uncertainty (line and yellow shaded region) 
are all taken from the survey of 
O`Sullivan et al. (2003). 
The blue 
(blue filled squares) and red (red open squares) globular cluster
density profiles are shown along with a beta model fits (blue and
red dotted lines). The galaxy surface brightness profile is shown
by a dashed line.
All profiles
  have an arbitrary vertical normalisation. For most galaxies,
  the outer slopes of the X-ray and blue GC profiles are similar,
  as are the galaxy surface brightness and red GC profiles.
}
\end{center}
\end{figure*}



\section*{Acknowledgments}

We would like to thank A. Romanowsky, R. Crain, J. Strader,
C. Frenk and G. Poole for useful
discussions.  Thanks to V. Pota, K. Woodley, K. Rhode, L. Spitler
and S. Kartha 
for supplying globular cluster profile data.  
EOS acknowledges support from the 
European Community under the Marie Curie
Research Training Network, and from the National Aeronautics and Space
Administration through Chandra Award Number AR1-12014X issued by the
Chandra X-ray Observatory Center, which is operated by the Smithsonian
Astrophysical Observatory for and on behalf of the National Aeronautics
and Space Administration under contract NAS8-03060. 
Some data used in this research comes from the Hyperleda and NED databases.
We also thank the referee for their careful reading and useful
comments.

\end{document}